# NMR Simulation of an Eight-State Quantum System

A. K. Khitrin and B. M. Fung

*Department of Chemistry and Biochemistry, University of Oklahoma,*

*Norman, Oklahoma 73019-0370*

**Abstract**

The propagation of excitation along a one-dimensional chain of atoms is simulated by means of NMR. The physical system used as an analog quantum computer is a nucleus of $^{133}$Cs (spin 7/2) in a liquid crystalline matrix. The Hamiltonian of migration is simulated by using a special 7-frequency pulse, and the dynamics is monitored by following the transfer of population from one of the 8 spin energy levels to the others.

## 1. Introduction

As it was explained by Feynman [1], classic computers are not efficient in simulating quantum physics. The reason is the large dimensionality of corresponding Hilbert spaces, even for very simple quantum systems. As an example, accurate simulation of dynamics of only a dozen of coupled spins ½ is beyond the reach of modern computers. Thus, simulation of quantum objects was the first and natural task that stimulated attempts to build quantum computers. After the discovery of quantum algorithms [2-4], which are more efficient than classic ones, an intense search for systems that can perform as universal quantum computers started. At present, systems of interacting nuclear spins are among the best candidates: they can have long relaxation and decoherence times, can be manipulated in desired ways, and spin degrees of freedom are well isolated from other degrees of freedom (lattice). The idea of using pseudopure states instead of pure quantum states [5,6] opened a possibility of using liquid-state NMR for ensemble quantum computing. Since then, a large number of work has been published, where various aspects of quantum computing have been realized by means of NMR (see [7] for a recent review).

While the possibility of building universal quantum computers of practical importance is still a disputable item, the task of creating specialized or analog quantum computers seems much more feasible. A trivial reason for such optimism is that any quantum system simulates itself. At the same time, we consider that the goal of quantum simulations is not to copy any existing physical system, but rather, to reproduce the behavior of idealized model systems with well-defined Hamiltonians. The Hamiltonian of the physical system used for simulation should be as close as possible to that of the model, and flexibility in changing the parameters of the Hamiltonian is definitely a plus. Again, nuclear spins are good candidates for practical realizations, and modern NMR spectrometers give the possibility of necessary manipulations with a high degree of coherent control. The first results of quantum simulations by means of NMR have been presented recently [8,9]. The simulated model systems were a four-level truncated oscillator [8] and a three-spin effective Hamiltonian [9]. Here we present the results of simulation of quantum dynamics in a more realistic physical model that describes the migration of excitation along a linear chain of atoms.

2. **Simulated model system**

We will consider a model system with the Hamiltonian

$$\mathcal{H} = \lambda \sum_{n=0}^{6} (a^{+}_n a_{n+1} + a_n a^{+}_{n+1}), \qquad (1)$$

where $a_n$ and $a^+_n$ are the annihilation and creation operators, respectively, for the site $n$, and $\lambda$ is the interaction constant. This Hamiltonian describes the migration of excitation or particle of any physical nature in an eight-site linear chain. It conserves the total number of particles. Since we will model the motion of a single particle, the necessary Hilbert space is reduced to eight dimensions and the Hamiltonian can be written as

$$\mathcal{H} = \lambda \begin{pmatrix} 0 & 1 & 0 & 0 & 0 & 0 & 0 & 0 \\ 1 & 0 & 1 & 0 & 0 & 0 & 0 & 0 \\ 0 & 1 & 0 & 1 & 0 & 0 & 0 & 0 \\ 0 & 0 & 1 & 0 & 1 & 0 & 0 & 0 \\ 0 & 0 & 0 & 1 & 0 & 1 & 0 & 0 \\ 0 & 0 & 0 & 0 & 1 & 0 & 1 & 0 \\ 0 & 0 & 0 & 0 & 0 & 1 & 0 & 1 \\ 0 & 0 & 0 & 0 & 0 & 0 & 1 & 0 \end{pmatrix}. \qquad (2)$$

In what follows, we will use dimensionless time by setting $\lambda = 1$. The quantum state $|n\rangle$, $0 \leq n \leq 7$, corresponds to excitation localized at the $n$-th atom. The population of a particular state gives the probability of finding the excitation at this location. The initial condition $\rho(0) = |0\rangle\langle 0|$ corresponds to excitation initially localized at one end of the chain.

### 3. Physical system

A very convenient eight-level (three qubit) system is the nucleus of $^{133}$Cs ($I = 7/2$, 100% natural abundance) in an anisotropic environment [10]. It is realized by a 54% solution of cesium pentadecafluorooctanoate in $D_2O$ at 48 $^0$C. Considerable residual quadrupolar splitting (6.0 kHz) between neighboring peaks in equidistant NMR spectrum makes this system suitable for manipulations with selective pulses.

In the interaction representation, the Hamiltonian of interaction with radio-frequency (rf) magnetic field will have the form (2) if the system is simultaneously irradiated at all 7 single-quantum transition frequencies and the amplitudes of the rf harmonics are inversely proportional to the corresponding spin matrix elements <I,m−1|I_|I,m> = $\sqrt{(I+m)(I-m+1)}$.

Instead of creating the pure initial state ρ(0) = |0><0|, it is sufficient to prepare a pseudopure state [5,6] with only a deviation part of the density matrix having the desired form. A very efficient way to prepare such pseudopure state is to simultaneously irradiate six upper single-quantum transitions so that the corresponding transition matrix elements (including the spin matrix elements) are, respectively, about 0.7,0.9,1,1,0.9,0.7 [10]. Under such irradiation the populations of the 7 upper levels, starting with equilibrium distribution, begin to change and simultaneously cross at some point, while the population of the ground state remains unchanged. This point of crossing determines the duration of the rf pulse. The off-diagonal matrix elements can be averaged either by pulsed gradients of magnetic field [6,11] or by temporal averaging [12]. It was convenient to use both methods in our experiments.

The scheme of experiment is shown in Fig. 1. The first shaped pulse is a sum of 6 rf harmonics whose amplitudes are adjusted in a way described above, and has a duration of 600 μs. This pulse, followed by a 1 ms pulsed field gradient (5 G/cm) to average out the off-diagonal elements of the density matrix, creates the pseudopure ground state. The second shaped pulse, which is a sum of 7 rf harmonics, creates the necessary Hamiltonian for the evolution period. Its length was varied from 0 to 600 μs. The phase of this pulse was incremented by π/2 in each successive experiment with respect to the phase of the last pulse, which is a π/20 reading pulse. The total number of transients was set to a multiple of 4. This averaged out the off diagonal elements of the density matrix (and their contribution to the observed signal) after the evolution period. The small-angle reading pulse is chosen to satisfy a linear response condition and to give accurate differences of populations for neighboring pairs of states.

## 4. Results and discussion

The equilibrium $^{133}$Cs NMR spectrum is shown at the top of Fig.2. Bars on the right schematically represent the excess populations of the levels. At equilibrium, all the differences of populations are equal and the intensities of the peaks are proportional to squares of the spin matrix elements, which are 7:12:15:16:15:12:7. Because of relaxation and exchange effects, the linewidths of the seven peaks are different, ranging from 10 Hz for the central peak to 130 Hz for the outermost peaks. These differences are reduced in the displayed spectra by using a 100 Hz broadening factor for the Fourier transform. The spectrum in the middle shows the spectrum of the initial pseudopure ground state after an evolution time $\tau = 0.13$, and it is still very close to the initial state. At longer evolution times the excessive population propagates, in the form of a wave packet, to other states (sites). At $\tau = 1.3$, as seen in the bottom spectrum, deviations of populations for the three neighbor sites are developed.

After each length of the evolution period the populations can be obtained by integrating the spectra (an integral for each peak is proportional to the difference in the populations of corresponding pair of states). The results are presented in Fig. 3. The populations of only four of the states are shown. The excessive populations of the other four states remain small at the displayed time scale. The dimensionless time in Fig. 3 is proportional to the product of actual evolution time and the amplitude of rf harmonics in the evolution shaped pulse. The duration of this pulse is limited by the relaxation processes ($T_2$ for the outer peaks is less than 10 ms). It is possible to increase the simulated times by increasing the amplitudes of the rf harmonics, but it will lead to loss of selectivity (the harmonics will start exciting transitions at neighboring frequencies). The parameters used in our experiment was a compromise between the effect of relaxation and sufficient selectivity of the pulses.

To assess the validity of the results of the experiment, a theoretical calculation of the migration produced by the Hamiltonian (2) was made. Though this dynamic problem can be solved exactly even for a finite number of sites [13], it was more convenient for our small system to perform a numerical simulation. The results are shown by curves in Fig. 3 for comparison with our experimental results of quantum simulation. The comparison shows that, although the accuracy is quite moderate, our NMR system simulates the propagation of excitation in the idealized model system, up to the third neighbor along the chain, without substantial loss of quantum coherence. At the same time, due to noticeable spin-lattice relaxation, the life time of quantum coherence is insufficient to simulate longer evolution times, when the wave of excitation reaches the opposite end of the chain, reflects and moves back.

An amazing feature of the "quantum processor" used in this work is that it is not even an atomic size object, but a single nucleus, which is eight orders of magnitude smaller. Of coarse, there was about $3\times10^{20}$ nuclei in the sample, but this number was used only to produce a detectable signal and avoid statistical averaging, while the nontrivial dynamical problem was solved by quantum motion of a single nucleus in external electric and magnetic fields. Although the model system studied here is still very simple and can be readily simulated by classical computers, our results demonstrate the potential of NMR simulation and reveal some of its limitations.

## Acknowlegement

This work was supported by the National Science Foundation under grant numbers DMR-9700680 and DMR-9809555 (Ferroelectric Liquid Crystal Materials Research Center).

## References

[1] R.P. Feynman, Int. J. Theor. Phys. **21**, 467 (1982).

[2] D. Deutch and R. Jozsa, Proc. R. Soc. Lond. A, **439**, 553 (1992).

[3] P. Shor, Proc. 35$^{th}$ Ann. Symp. on Found. of Computer Science (IEEE Comp. Soc. Press, Los Alamitos, CA, 1994) 124.

[4] L.K. Grover, Phys. Rev. Lett. **79**, 325 (1997).

[5] N.A. Gershenfeld and I.L. Chuang, Science **275**, 350 (1997).

[6] D.G. Cory, A.F. Fahmy, and T.F. Havel, Proc. Natl. Acad. Sci. USA **94**, 1634 (1997).

[7] J.A. Jones, Progr. NMR Spectroscopy, in press.

[8] S. Somaroo, C.H. Tseng, T.F. Havel, R. Laflamme, and D.G. Cory, Phys. Rev. Lett. **82**, 5381 (1999).

[9] C.H. Tseng, S. Somaroo, Y. Sharf, E. Knill, R. Laflamme, T.F. Havel, and D.G. Cory, Phys. Rev. A **61**, 012302 (1999).

[10] A. Khitrin, H. Sun, and B.M. Fung, Phys. Rev. A **63,** 0203XX (2001).

[11] A.K. Khitrin and B.M. Fung, J. Chem. Phys. **112**, 6963 (2000).

[12] E. Knill, I. Chuang, and R. Laflamme, Phys. Rev. A **57**, 3348 (1998).

[13] E.B. Fel'dman, R. Bruschweiler, and R. Ernst, Chem. Phys. Lett. **294**, 297 (1998).

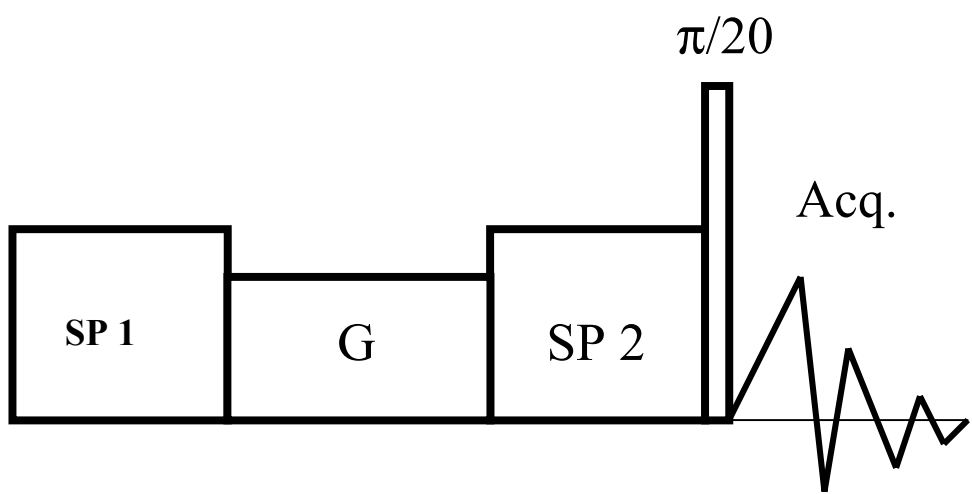

Fig. 1. The scheme of the NMR experiment. SP1 is a shaped pulse for preparing the pseudopure ground state, G is a field gradient pulse, SP2 is a shaped pulse for the evolution period, and $\pi/20$ is the reading pulse.

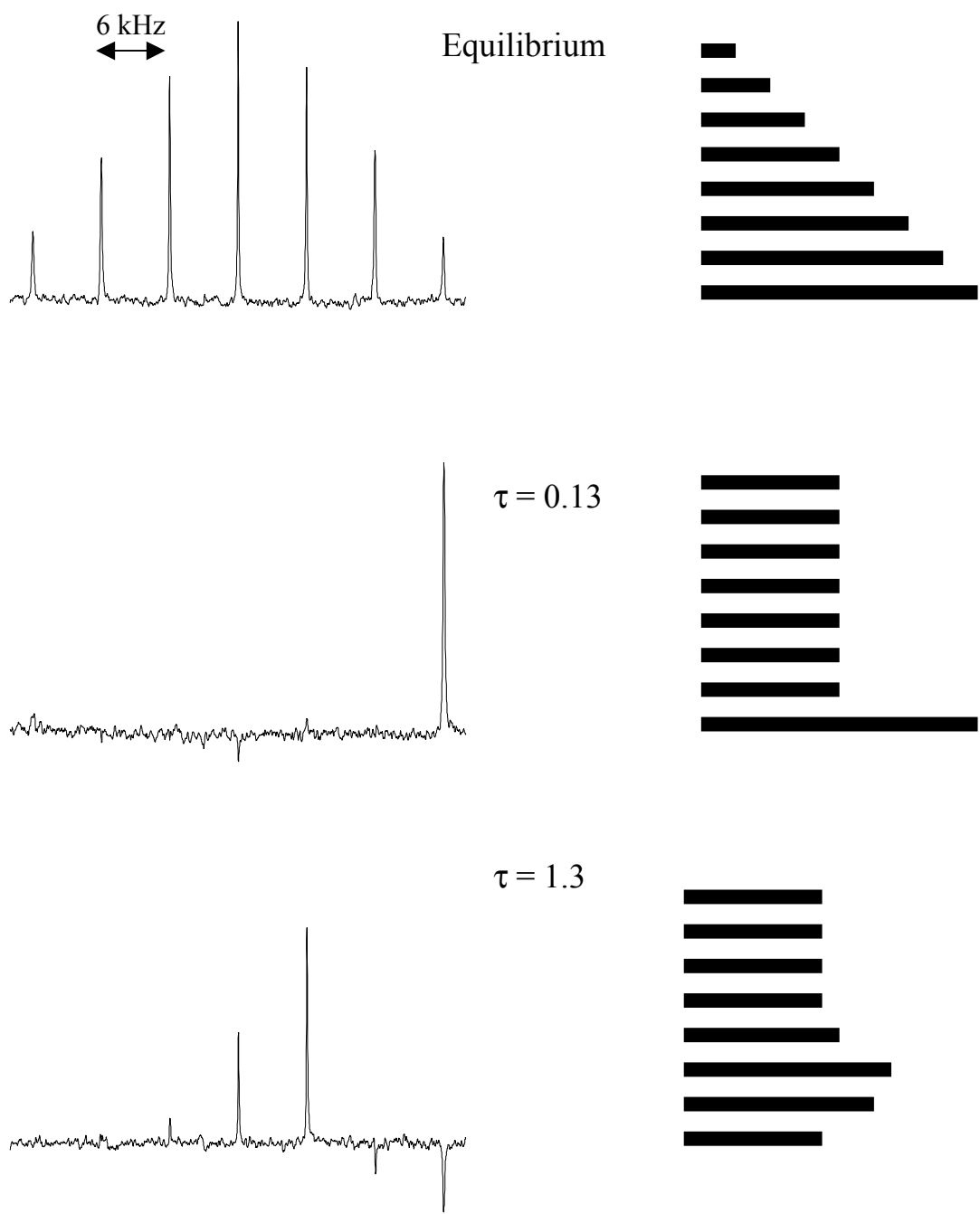

Fig. 2. $^{133}$Cs NMR equilibrium spectrum (top) and the spectra corresponding to two different evolution times. The bars on the right show excessive populations of the levels.

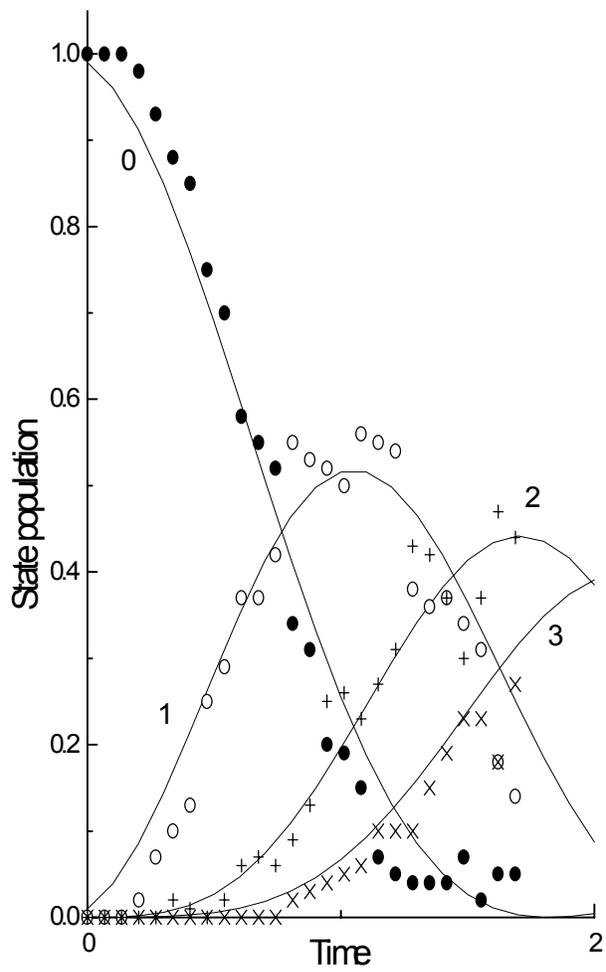

Fig. 3. Simulated populations of the four states (sites); the curves are theoretical results. "0" is the ground state, "1" is the first excited state, and so on.